\documentclass[onecolumn,amsmath,amssymb]{qm2008_abs}
\usepackage{graphicx}
\usepackage{dcolumn}
\usepackage{bm}
\graphicspath{{img/}}
\begin{document}

\title{{\Large Two-particle Azimuthal Correlations of High-$p_T$ Charged Hadrons at the CERN SPS}}

\bigskip
\bigskip
\author{\large Marek SZUBA for the NA49 Collaboration}
\email{Marek.Szuba@cern.ch}
\affiliation{Faculty of Physics, Warsaw University of Technology, Warszawa, Poland}
\bigskip
\bigskip

\begin{abstract}
  \leftskip1.0cm
  \rightskip1.0cm
  Two-particle azimuthal correlations of high-$p_T$ hadrons can serve as a probe of interactions
  of partons with the dense medium produced in high-energy heavy-ion collisions. First NA49 results
  on such correlations are presented for central and mid-central Pb+Pb collisions at 158$A$~GeV beam
  energy, for different centrality bins and charge combinations of trigger and associate particles.
  These results feature a flattened away-side peak in the most central collisions, which is
  consistent with expectations of the medium-interaction scenario. A comparison with CERES Pb+Au
  results at the same energy, as well as with PHENIX Au+Au results at the top RHIC energy, is provided.
\end{abstract}

\maketitle

\section{Introduction}

It is believed that one of the signatures of a hot, dense medium expected to appear
in most central high-energy collisions of heavy ions, will be modification of properties
of jets --- highly collimated streams of particles, originating from hard scattering of partons
and therefore produced early in a collision --- as a result of their interaction with that
medium~\cite{WangGyulassy}. Unfortunately, direct reconstruction of jets in the presence
of a large background typical of such events is an extremely challenging task. One possible answer
to this problem is studying two-particle azimuthal correlations, which allow one to extract
the jet signal from the soft background by taking advantage of the fact jet particles are strongly
correlated in azimuth. In recent years RHIC experiments have made this approach highly successful
by not only observing the expected signatures, but also providing, through further investigations,
evidence which has led to a major change of expectations regarding properties of the hot medium
(the ``perfect liquid'' description; see \textit{e.g.}~\cite{STARresults,PHENIXresults}). However,
until recently (\cite{CERESfirst}) no analyses of this sort were performed in the energy range
of the CERN SPS.

\section{Experimental Setup}

The NA49 detector is a large-acceptance spectrometer dedicated to the study of hadron production
in fixed-target nucleon-nucleon, nucleon-nucleus and nucleus-nucleus collisions at a wide range
of energies offered by the CERN SPS~\cite{na49nim}. During its eight years of running it was
used to register Pb+Pb collisions at 20$A$, 30$A$, 40$A$, 80$A$ and 158$A$ GeV, C+C and Si+Si
collisions at 40$A$ and 158$A$ GeV, along with p+p and p+Pb interactions at 158$A$ GeV.

The main components of the detector are four time projection chambers, used for tracking as well
as particle identification by dE/dx. Two of the chambers, the so-called Vertex TPCs, are located inside
two superconducting magnets (with field intensity of 1.5 and 1.1~T, respectively) positioned along
the beam axis right downstream of the target, with the other two (``Main TPCs'') placed further
downstream on both sides of the beam line. Two Time of Flight walls located beyond the MTPCs complement
particle identification in the intersection region of Bethe-Bloch dE/dx bands. Centrality selection in
nucleus-nucleus collisions is based on the projectile spectator energy deposited in the Veto Calorimeter,
located at the downstream end of the experiment.

The present analysis is based on 3~million central Pb+Pb collisions at 158$A$~GeV recorded by NA49
in the year 2000.

\section{Data Analysis}

The two-particle correlation function is calculated independently in four event-centrality
bins: 0-5~\%, 5-10~\%, 10-15~\%, and 15-20~\%. The transverse momentum bins are $2.5~GeV/c \le p_T^{trg} \le 4.0~GeV/c$
for trigger particles and $1.0~GeV/c \le p_T^{trg} \le 2.5~GeV/c$ for associates. 

\begin{figure}[htbp]
  \begin{center}
    \includegraphics[width=\textwidth]{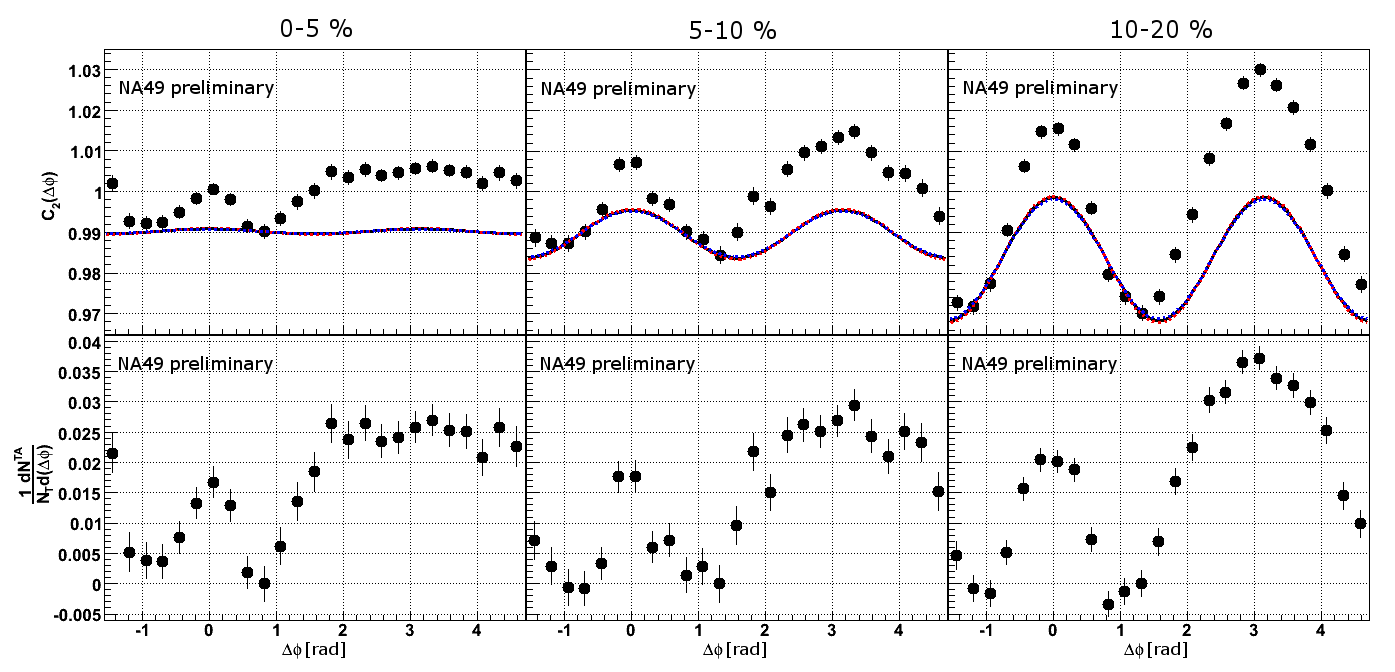}
  \end{center}
  \caption{\textbf{Top}: Two-particle azimuthal correlation functions of charged hadrons in Pb+Pb collisions
  at 158$A$~GeV, for centrality bins 0-5~\% (left), 5-10~\% (middle) and 10-20~\% (right). The solid
  lines illustrate ZYAM-normalised flow contribution to the function, with the dashed ones
  indicating modulation due to statistical uncertainties on flow coefficients.
  \textbf{Bottom}: per-trigger conditional yield of associate particles obtained by normalising the
  flow-subtracted correlation function, again in three centrality bins. All errors are statistical only.}
  \label{fig:c2phi_yield}
\end{figure}

Following the prescription of the PHENIX Collaboration~\cite{phenixMethod} we define the correlation
function as a ratio of two normalised distributions of $\Delta\phi = \phi_{asc} - \phi_{trg}$, where
$\phi$ is the azimuthal angle: $N_{corr}(\Delta\phi)$, in which both particles come from the same event,
and $N_{mix}(\Delta\phi)$, in which the trigger and the associate originate from different events.
Division by the mixed-event spectrum accounts for non-uniform detector acceptance.
\begin{equation}
  C_{2}(\Delta\phi) = \frac{N_{corr}(\Delta\phi)}{N_{mix}(\Delta\phi)}
  \frac{\int{N_{mix}(\Delta\phi')}\mathrm{d}(\Delta\phi')}{\int{N_{corr}(\Delta\phi')}\mathrm{d}(\Delta\phi')}
  \label{eqn:c2phi}
\end{equation}
In this analysis the mixing is accomplished with a sliding window of up to 50 most recent events in each
centrality bin. The resulting correlation functions are shown in the top row of Fig.~\ref{fig:c2phi_yield}.

Although the correlation function as defined in Eq.~\ref{eqn:c2phi} is free of acceptance effects,
it still includes all physical correlations: conservation laws, resonance decays, flow, quantum statistics,
parton fragmentation \textit{etc.} The \emph{two-source model}~\cite{twosource}, a widely-used description
of azimuthal correlations in nucleus-nucleus collisions, postulates that the function $C_2$ can be decomposed
into only hard-scattering and flow contributions:
$C_2(\Delta\phi) = C_2^{jet}(\Delta\phi) + a \left[ 1 + 2 \langle v_2^T \rangle \langle v_2^A \rangle cos(2\Delta\phi) \right]$.
The two flow terms $v_2^T$ and $v_2^A$ have been obtained from an independent reaction-plane analysis~\cite{flowMethods},
see Table~\ref{tbl:v2values} for the values. Finally, we obtain the factor $a$ by employing the Zero Yield
At Minimum (ZYAM) assumption~\cite{twosource}. The flow contribution to the correlation function is
illustrated in the top row of Fig.~\ref{fig:c2phi_yield} by the solid line, with the dashed ones
indicating modulation due to statistical uncertainties of $\langle v_2^T \rangle$ and $\langle v_2^A \rangle$.

\begin{table}[htb]
  \begin{tabular} { | r | c | c | }
    \hline
    Centrality & $v_2^T$ & $v_2^A$ \\ \hline
    0-5 \% & 0.022 $\pm$ 0.003 & 0.01395 $\pm$ 0.00027 \\ \hline
    5-10 \% & 0.073 $\pm$ 0.003 & 0.04079 $\pm$ 0.00029 \\ \hline
    10-20 \% & 0.117 $\pm$ 0.003 & 0.06570 $\pm$ 0.00026 \\ \hline
  \end{tabular}
  \caption{Flow coefficients used in the analysis. Only statistical errors are listed.}
  \label{tbl:v2values}
\end{table}

Having subtracted the flow one can use the hard-scattering component of the correlation function
to calculate per-trigger conditional yield of associate particles:
\begin{equation}
  \hat{J}(\Delta\phi) = \frac{1}{N_T} \frac{\mathrm{d}N^{TA}}{\mathrm{d}\Delta\phi} =
  \frac{C_2^{jet}(\Delta\phi)}{\int{C_2(\Delta\phi')}\mathrm{d}(\Delta\phi')} \frac{N^{TA}}{N_{T}},
  \label{eqn:c2yield}
\end{equation}
where $N_T$ is the number of trigger particles, and $N^{TA}$ --- the number of same-event
trigger-associate pairs. Yields obtained this way are shown in the bottom row of Fig.~\ref{fig:c2phi_yield}.

\begin{figure}[htbp]
  \begin{center}
    \includegraphics[width=\textwidth]{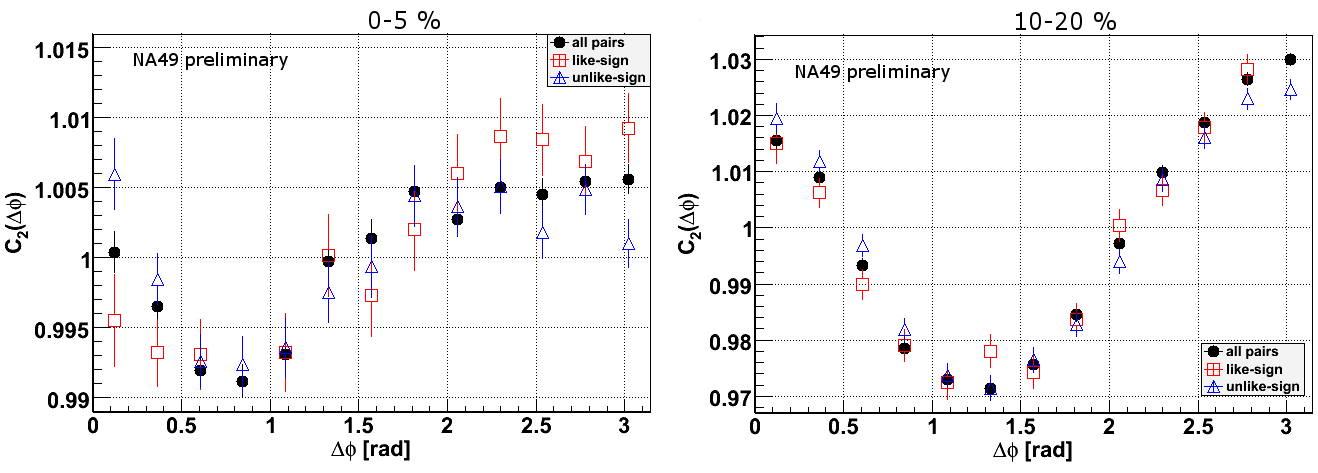}
  \end{center}
  \caption{Two-particle azimuthal correlation functions for different combinations of electric charge
  of trigger and associate particles: like-sign pairs (open squares), unlike-sign pairs (open
  triangles) and no constraints (full circles).}
  \label{fig:chargeSelection}
\end{figure}

Next, Fig.~\ref{fig:chargeSelection} shows correlation functions obtained by imposing additional constraints
on the electric charge of trigger and associate particles. Like- and unlike-sign pairs have been
considered separately and compared to the correlation functions without charge selection, for centrality
bins 0-5~\% and 10-20~\%.

\begin{figure}[htbp]
  \begin{center}
    \includegraphics[width=0.9\textwidth]{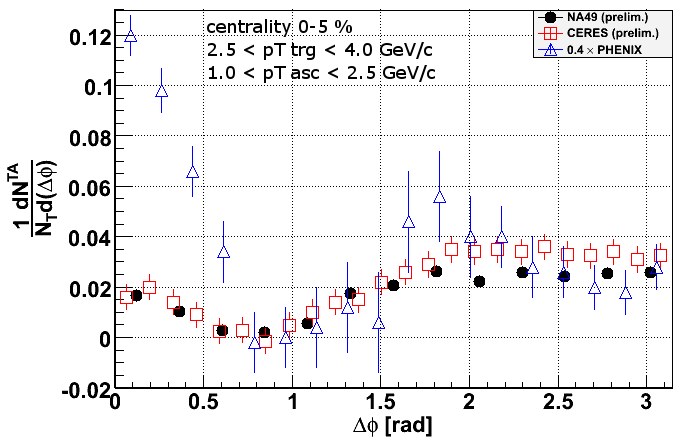}
  \end{center}
  \caption{Per-trigger conditional yield for most central ($\sigma/\sigma_{geom} = 0-5~\%$) nucleus-nucleus
  collisions. Full circles: Pb+Pb at 158$A$~GeV (NA49 preliminary). Open squares:
  Pb+Au at 158$A$~GeV (CERES preliminary). Open triangles: Au+Au at $\sqrt{s_{NN}}$~=~200~GeV (PHENIX),
  scaled down to match SPS results around the minimum.}
  \label{fig:expComparison}
\end{figure}

Finally, Fig.~\ref{fig:expComparison} compares the conditional yield of NA49 with that from Pb+Au
collisions at 158$A$~GeV as observed by the CERES experiment at the SPS~\cite{ceres} and that from
Au+Au collisions at $\sqrt{s_{NN}}$~=~200~GeV as observed by the PHENIX experiment at the RHIC~\cite{phenixMethod}.
In order to facilitate comparison of the shape of the functions, PHENIX results have been scaled down by the factor
0.4 to match the amplitude of SPS yields near the minimum. All three yields have been obtained for the same centrality
bin as well as trigger and associate particle $p_T$ ranges.

\section{Results and Summary}

\noindent

NA49 has measured two-particle azimuthal correlation functions in central and mid-central Pb+Pb collisions
at 158$A$~GeV. The away side of the correlation function in the most central collisions shows plateau-like
structure, visible even before subtraction of flow and significantly different from the away-side shape
in more peripheral bins. Such behaviour is consistent with qualitative expectations for a hot, dense medium
in the most central high-energy nucleus-nucleus interactions. Moreover, NA49 results agree with those from
the CERES experiment at the SPS and are in qualitative agreement with RHIC results. Finally, the observed
difference in amplitude of the near-side peak of the correlation function of like- and unlike-sign pairs,
is consistent with local charge conservation which could be associated with fragmentation of partons.

\end{document}